\pdfoutput=1
\documentclass[11pt,twoside,a4paper,cmspaper,final,collab]{cms-tdr}
\def\svnVersion{06f8e91}\def\svnDate{2025/04/04}\def\cmsCernNoTag{CERN-EP-2024-002}\def\cmsCernDate{\today}\def\cmsMessage{Published in Physical Review Letters as \href{http://dx.doi.org/10.1103/PhysRevLett.134.131802}{\doi{10.1103/PhysRevLett.134.131802}.}}
\begin{document}\cmsNoteHeader{EXO-19-006}

\cmsNoteHeader{EXO-19-006}

\newcommand{\dEdx}{\ddinline{E}{x}\xspace}
\newcommand{\partialslash}{\ensuremath{\partial\hspace{-0.5em}/}\xspace}
\newcommand{\Aslash}{\ensuremath{A\hspace{-0.65em}/}\xspace}
\newcommand{\Bslash}{\ensuremath{B\hspace{-0.65em}/}\xspace}
\newcommand{\nLow}{\ensuremath{N_{\text{hits}}^{\text{low \dEdx}}}\xspace}
\newcommand{\nLowQSq}{\ensuremath{N_{\text{hits}, Q^{2}}^{\text{low dE/dx}}}\xspace}
\newlength\cmsTabSkip\setlength{\cmsTabSkip}{1ex}
\newlength\cmsFigWidth\ifthenelse{\boolean{cms@external}}{\setlength\cmsFigWidth{\columnwidth}}{\setlength\cmsFigWidth{0.6\textwidth}}

\ifthenelse{\boolean{cms@external}}{\providecommand{\suppMaterial}{the supplemental material~\cite{supp}}}{\providecommand{\suppMaterial}{Appendix~\ref{app:suppMat}\nocite{supp}\xspace}}

\title{Search for fractionally charged particles in proton-proton collisions at \texorpdfstring{$\sqrt{s} = 13\TeV$}{sqrt(s) = 13 TeV}}

\author*[inst1]{David Vannerom}

\date{\today}

\abstract{
A search is presented for fractionally charged particles with charge below $1e$, using their small energy loss in the tracking detector as a key variable to observe a signal. The analyzed data set corresponds to an integrated luminosity of 138\fbinv of proton-proton collisions collected at $\sqrt{s} = 13\TeV$ in 2016--2018 at the CERN LHC. This is the first search at the LHC for new particles with a charge between $e/3$ and $0.9e$, including an extension of previous results at a charge of $2e/3$. Masses up to 640\GeV and charges as low as $e/3$ are excluded at 95\% confidence level. These are the most stringent limits to date for the considered Drell-Yan-like production mode.
}

\hypersetup{%
pdfauthor={CMS Collaboration},%
pdftitle={Search for fractionally charged particles in pp collisions at sqrt(s)=13TeV},%
pdfsubject={CMS},%
pdfkeywords={CMS, exotica, fractionally charged particles}}

\maketitle

Colorless freely-propagating particles with a fractional electric charge, or fractionally charged particles (FCPs), have never been observed.
In particular, such particles with a charge lower than the electron charge $e$ remain unseen, although nothing from first principle prevents their existence.
The theoretical motivations for the existence of FCPs are closely related to possible hidden symmetries and fields weakly or not connected to the standard model (SM) fields.
For example, these particles are predicted in the hypercharge portal scenario, where a new U(1) symmetry is added to the SM Lagrangian, and the corresponding new boson, called a dark photon, kinetically mixes with the SM hypercharge field~\cite{Holdom:1985ag}.
New U(1) symmetries are introduced, e.g., in string theory phenomenology~\cite{Abel:2008ai,WEN1985651}, and in Grand Unified Theories or reheating cosmology~\cite{Gan:2023jbs}, as remnants of large symmetries that break down to the SM symmetry group and extra U(1) groups~\cite{Langacker:2008yv}.
They also appear in theories of quantum gravity~\cite{Shiu:2013wxa}.
Breaking supersymmetry in a hidden sector is another way to generate such new Abelian groups~\cite{Morrissey:2009ur}.

\begin{figure}[!htb]
  \centering
  \includegraphics[width=0.75\cmsFigWidth]{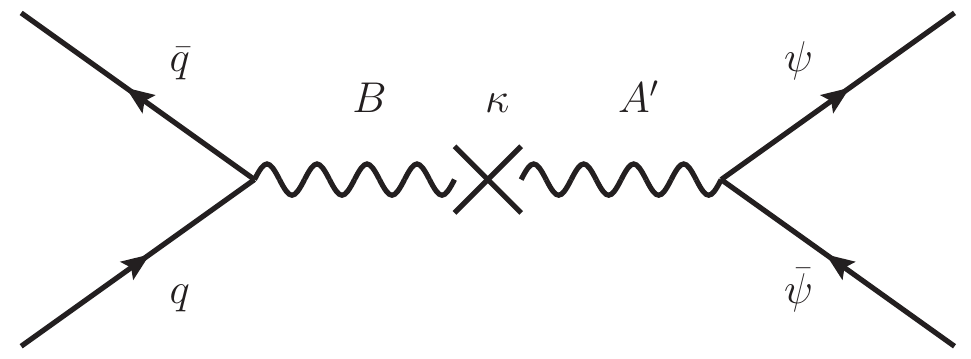}
  \caption{Feynman diagram of the kinetic mixing between the SM hypercharge field $B$ and a new U(1) gauge field $A'$. In this work, FCPs are considered to be pair-produced through this process.\label{fig:feynDiag}}
\end{figure}

When adding to the SM Lagrangian such a U(1) gauge field $A'$ that couples to a new Dirac fermion $\psi$ with mass $M$ through a coupling constant $e'$, the kinetic mixing between the new gauge field and the SM hypercharge field $B$ leads (after a field redefinition ${A'_{\mu} \to A'_{\mu} - \kappa B_{\mu}}$) to the $\psi$ field acquiring an effective nonzero weak hypercharge $e'\kappa$ and thus a coupling to the SM photon and \PZ boson after electroweak symmetry breaking:
\begin{equation*}
\mathcal{L} \supset -\frac{1}{4}A'_{\mu\nu}A'^{\mu\nu}+i\bar{\psi}(\partialslash+ie'\Aslash'-ie'\kappa\Bslash+iM)\psi.
\end{equation*}
Figure~\ref{fig:feynDiag} illustrates this kinetic mixing process.
The electric charge ${Q = e'\kappa\cos\theta_{\PW} \equiv \epsilon e}$ of the new fermion $\psi$, where $\theta_{\PW}$ denotes the weak mixing angle, can have any value.
For the purpose of this search, the sign of the charge is irrelevant and we will always use $Q$ as a positive value.
In the SM, fermions couple to electroweak vector bosons through vector and axial-vector couplings ${g_{\mathrm{V}}=I_{3}-2Q \sin^{2}\theta_{\PW}}$ and ${g_{\mathrm{A}}=I_{3}}$.
Since $\psi$ only carries U(1) charge, it has no SU(2) weak isospin $I_{3}$, and we assume the same value for $\theta_{\PW}$ as in the SM.
The matter field $\psi$ is what we refer to as the FCP.
We consider FCPs to be pair-produced in a Drell--Yan (DY)-like process with an associated DY cross section where the charge and mass of the final state fermion are taken to be that of the FCP.

In this Letter, we present a search for FCPs in proton-proton ($\Pp\Pp$) collisions at a center-of-mass energy of 13\TeV with the CMS detector at the CERN LHC.
The data analyzed correspond to an integrated luminosity of 138\fbinv collected in 2016--2018.
For FCP charges lower than $e$, the FCP-matter interaction cross section is smaller than that of unit charge particles such as muons, resulting in an extended interaction length and a smaller energy loss per unit length \dEdx.
The strategy adopted in this search is to measure the ionization of particles passing through the tracking detector to discriminate signal from background events.
A similar search was pursued previously by CMS using $\Pp\Pp$ collision data corresponding to an integrated luminosity of 5.0\fbinv at ${\sqrt{s} = 7\TeV}$~\cite{CMS:2012xi}.
Another CMS analysis, also based on ionization measurements but using a different approach~\cite{CMS:2013czn}, combined data sets of 5.0\fbinv at 7\TeV and 18.8\fbinv at 8\TeV.
These results exclude at 95\% confidence level (\CL) the existence of FCPs with a charge of $2e/3$ in the mass range 100--460\GeV.
Exclusions for a charge of $e/3$ have been recently understood to be affected by an issue with the muon trigger simulation at low charge, and have been updated (see Errata of Refs.~\cite{CMS:2012xi,CMS:2013czn}). In this work, we improve the simulation of the FCP signal in terms of \dEdx in the CMS muon detectors. This allows us to produce an exclusion result for a charge of $e/3$ and to extend the reach of the search for other probed FCP charge and mass scenarios.

At higher masses, the most stringent constraints are derived indirectly from the anisotropy of the cosmic microwave background, which can be affected by FCPs in the hot primordial plasma~\cite{Dubovsky:2003yn}. In models with a massless dark photon with coupling $e' = 0.1$, such constraints exclude the existence of FCPs with a mass larger than about 241\GeV~\cite{Vogel_2014}, for all charges considered in this paper.
These results are model-dependent and are complementary to those of the model-independent direct-detection search we pursue in this work.
At lower masses, FCPs have been searched for at colliders.
In particular, LEP constraints exclude the existence of particles with a charge larger than $2e/3$ and a mass below 95\GeV~\cite{Davidson:2000hf,OPAL:2003zpa}.
Future proposed dedicated experiments at the LHC intend to probe them down to $10^{-3}e$ for masses below half the $\PZ$ boson mass~\cite{Haas:2014dda,Ball:2016zrp,Kalliokoski:2023cgw}.
Among those, the milliQan experiment has already set limits in this region of the phase space~\cite{Ball:2020dnx} and is currently taking data with an improved detector~\cite{milliQan:2021lne}.
Through the reinterpretation of cosmic ray measurements, other bounds have been derived~\cite{Hu:2016xas,Plestid:2020kdm} that depend on the nature of dark matter and on FCP cosmic acceleration.
The range of these bounds overlaps with the phase space accessible at the LHC, with a lower edge as low as $10^{-4}e$.
Finally, two bulk matter searches using a Millikan oil drop method have set upper limits for FCP abundance in silicon oil ($1.17\times10^{-22}$ particles per nucleon for charges between $0.18e$ and $0.82e$)~\cite{PhysRevD.66.012002} and in mineral oil ($1.9\times10^{-23}$ particles per nucleon) and meteoritic material ($1.3\times10^{-21}$ particles per nucleon) for charges between $0.25e$ and $0.75e$~\cite{Kim:2007zzs}.

The central feature of the CMS apparatus is a superconducting solenoid of $6\unit{m}$ internal diameter, providing a magnetic field of $3.8\unit{T}$.
Within the solenoid volume are a silicon pixel and strip tracker, a lead tungstate crystal electromagnetic calorimeter, and a brass and scintillator hadron calorimeter, each composed of a barrel and two endcap sections.
The silicon tracker consists of pixel and strip modules that measure ionization clusters (hits) and the associated \dEdx arising from tracks in the pseudorapidity range $\abs{\eta} < 2.5$.
Between the 2016 and 2017 data-taking periods, the pixel tracker was fully replaced, adding an extra detection layer both in the central and forward sections of the detector~\cite{CMSTrackerGroup:2020edz}.
Muons are detected in gas-ionization chambers embedded in the steel flux-return yoke outside the solenoid.
They are measured in the range $\abs{\eta} < 2.4$, with detection planes made using three technologies: drift tubes, cathode strip chambers, and resistive-plate chambers.
A detailed description of the CMS detector, together with a definition of the coordinate system used and the relevant kinematic variables, can be found in Ref.~\cite{Chatrchyan:2008zzk}.

Isolated particles with transverse momentum ${\pt = 100\GeV}$ emitted with ${\abs{\eta} < 1.4}$ have track \pt resolution of 2.8\%.
In this $\eta$ range, the distribution of the number of tracker layers a particle sees is narrowly centered around 13 (14) with the old (new) pixel detector. 
The efficiency to reconstruct and identify muons is greater than 96\% for muons with \pt above 20\GeV~\cite{Sirunyan:2018}.
Matching muons to tracks measured in the silicon tracker results, for muons with \pt up to 100\GeV, in a relative \pt resolution of 1\% in the barrel and 3\% in the endcaps.
For higher-\pt muons, up to a \pt of 1\TeV, the resolution in the barrel remains better than 7\%.

The experimental signature of an FCP in CMS is very similar to that of a muon, i.e., a minimum-ionizing long-lived particle.
It is expected to leave a track both in the silicon tracker and in the muon chambers.
In order to assign a \pt to charged particles with curved trajectories in the CMS magnetic field, the track reconstruction algorithm assumes that particles carry an electric charge of $1e$.
Because their charge $Q$ is lower than $1e$, the measured \pt of FCPs is overestimated by a factor of $1/Q$.
This leads to an improved efficiency for the selection of high-\pt signal tracks.
On the other hand, FCPs ionize in the detector with a \dEdx proportional to $Q^2$.
As a consequence, FCPs produce a softer \dEdx spectrum than SM charged particles.
For smaller FCP charges ($Q < 0.5e$), energy deposits in the tracker and in the muon chambers have an increased probability to fall below the readout threshold, which may lead to a loss of signal efficiency because of tracks not being reconstructed.

Signal events are simulated using \PYTHIA~8.215~\cite{Sjostrand:2014zea} at leading order (LO).
Background events from ${\PW \to \PGm\PGn}$, ${\PZ \to \PGmp \PGmm}$, and \ttbar production are generated with \MGvATNLO~\cite{Alwall:2014hca} (v5\_2.2.2) at LO or next-to-LO (NLO), depending on the process. Diboson events are generated with \PYTHIA at LO.
All samples are generated with the corresponding 2016, 2017, and 2018 detector conditions.
The LO NNPDF3.0 PDF set~\cite{Ball:2017nwa} is used for 2016 samples, while the next-to-NLO NNPDF3.1 PDF set~\cite{Ball:2017nwa} is used for 2017 and 2018 samples.
Parton showering and fragmentation is performed for the 2016 event samples using the CUETP8M1 tune~\cite{Khachatryan:2015pea}, except for \ttbar events, for which the CUETP8M2T4 tune is used~\cite{CMS-PAS-TOP-16-021}.
For all 2017--2018 samples, the CP5 tune is used~\cite{Sirunyan:2019dfx}.
All generated events are passed through a simulation of the CMS detector, using \GEANTfour~\cite{Agostinelli:2002hh}, and its readout electronics and electronic signals from additional $\Pp\Pp$ interactions within the same or adjacent bunch crossings are overlaid.
In the muon chambers, FCP-matter interactions are generated using a combination of \GEANTfour and a parametrized framework to simulate the creation of individual electrons through ionization and their drift to the collecting electrodes.

Events of interest are selected using a two-tiered trigger system~\cite{Khachatryan:2016bia}.
In this search, we employ a single-muon trigger with a \pt threshold of 50\GeV~\cite{Sirunyan:2018}.
The muon-like particle reconstructed offline is required to have \pt above 55\GeV, \ie above the peak in the \pt spectrum of muons from \PW and \PZ boson decays.
Events are further selected if they contain at least one central, isolated, high-\pt track geometrically matched to a reconstructed muon-like particle.
The track must have ${\abs{\eta}< 1.5}$, and come from the primary vertex of the event.
To ensure a sufficient track quality, at least 5 hits are required overall in the tracker, among which at least one should be in the pixel detector.
For each track, the muon system is used to estimate the time at which the particle traveling at nearly the speed of light would have emerged from the primary vertex.
By convention, this time is redefined to have its mean centered around zero.
For ultrarelativistic particles, the resolution is of the order of 1.6\unit{ns}~\cite{Sirunyan:2018}.
As FCPs are massive and thus travel slower and arrive later in the muon chambers, tracks are selected if this time is positive.
This rejects about half of the muon background from SM processes.
Furthermore, the maximum angle between the track and any other track or muon-like particle with ${\pt > 35\GeV}$ is required to be smaller than 2.8 radians to reject back-to-back tracks from traversing cosmic ray muons.
These selection criteria define candidate tracks, to which the signal selection is then applied.
Events are retained for analysis if they contain 1 or 2 candidate tracks.

Because of their smaller ionization, FCP tracks have a larger number of tracker hits with low \dEdx than expected for particles with charge $e$.
To discriminate FCPs (signal) from muons (background), we count the number of track hits with a low value of \dEdx (\nLow).
The choice of the \dEdx threshold below which a hit is considered low is detailed later together with the background estimation.

Some instrumental effects that reduce the \dEdx values are not well simulated.
Radiation damage affects the ability for the tracker sensors to collect the ionization charge, leading to an increased probability for tracks to have a high number of low-\dEdx hits.
In particular, we observe the pixel detector layer closest to the beam to be strongly affected by radiation.
For this reason, and because the readout of this layer was defective in some data-taking periods~\cite{CMS:2021ime}, we choose not to count hits on this layer.
Furthermore, we observe the number of low-\dEdx hits to be underestimated in simulations, even for tracker layers further away from the collision point.
Therefore, the signal \dEdx spectrum in each of the tracker layers is corrected independently, using correction factors derived from comparisons of candidate tracks between data and simulations.
The effect of these corrections is to reduce the \dEdx value of a fraction of the hits.
These corrections are particularly important for data collected in the first half of 2016, when an electronic saturation problem in the strip tracker led to an increase in hits with low-\dEdx values~\cite{CMS-DP-2020-045,Butz:2018dum}.
Another experimental effect not well simulated is related to the edges of the tracker modules.
When a particle deposits energy close to the edge of a module, part of the charge is not collected, leading to a low-charge hit.
Because of the way the modules are physically arranged, this increases the probability of such tracks having a high multiplicity of low-\dEdx hits.
Consequently, we ignore tracks containing at least one hit close to the edge of a tracker module.
This selection has an efficiency of 70\% for events passing all other selection criteria, which is one of the largest inefficiencies for signal events passing the trigger criteria.
Finally, we observe an anomalous accumulation of hits with $\dEdx < 0.25\MeV/\cm$ for a few modules in the strip tracker with poor gain calibration, which we ignore.

We divide the total data set into four periods: early 2016 (19.9\fbinv), late 2016 (16.4\fbinv), 2017 (41.5\fbinv), and 2018 (59.7\fbinv).
Each period includes different instrumental effects.
Thus, the expected number of tracks from background and signal events is estimated separately in each of these four periods.
In the following, we demonstrate some features of the search using the 2018 data set, which is the largest and contributes most to the overall sensitivity.
The final results derive from a combination of all four data-taking periods.

\begin{figure}[!htb]
  \centering
  \includegraphics[width=\cmsFigWidth]{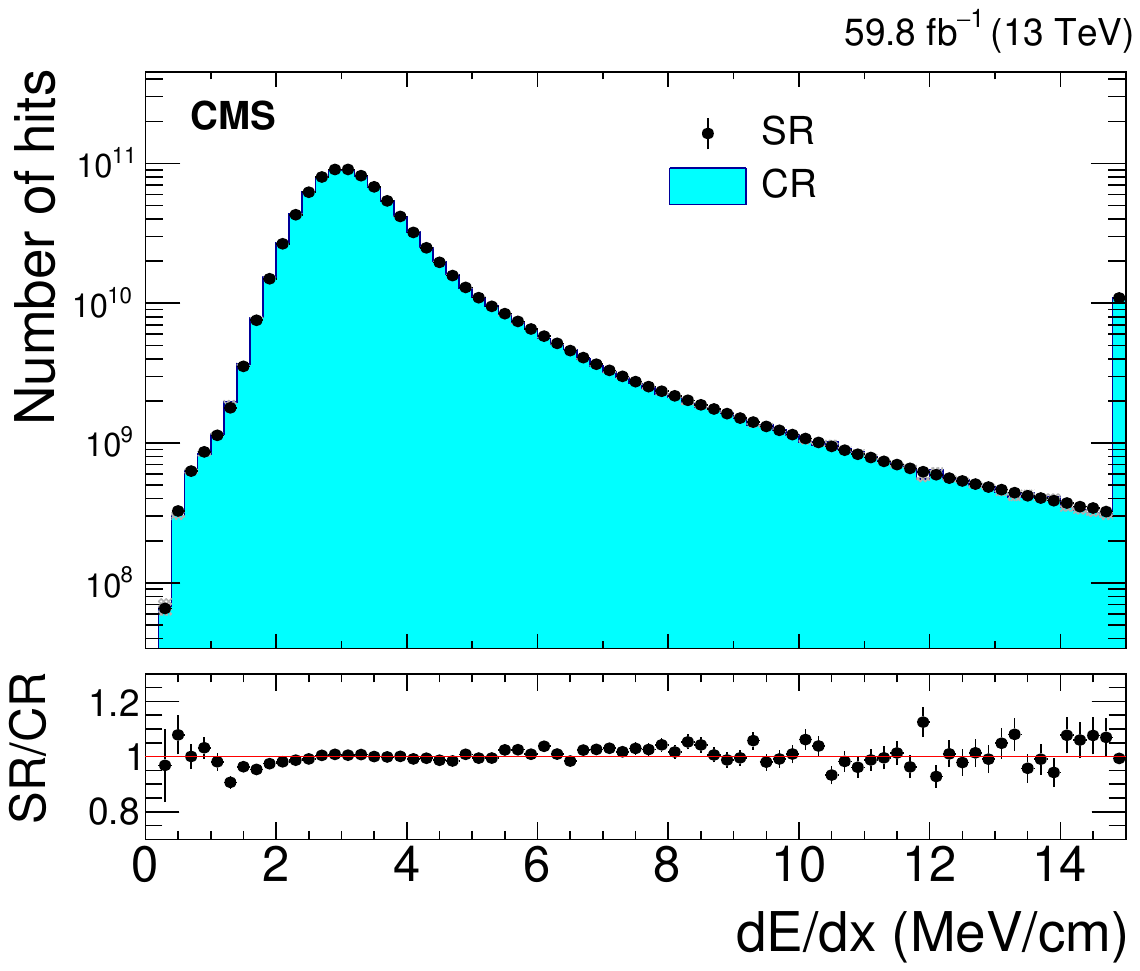}
  \caption{The \dEdx distribution for hits on candidate tracks in the CR and SR in 2018 data. The last bin is an overflow bin. The vertical bars and the shaded area correspond to the statistical uncertainty in the SR and the CR, respectively.}\label{fig:dEdx}
\end{figure}

We split our selected events into a search region (SR) and a control region (CR).
The SR consists of events with only one track and events with two tracks with an invariant mass smaller than 80 or larger than 100\GeV, where the mass assigned to the track is the muon mass.
The CR consists of events with two tracks that have an invariant mass in the range 80--100\GeV, and is dominated by \PZ boson production.
The CR has fewer tracks than the SR and is normalized to the number of tracks in the SR.
As shown in Fig.~\ref{fig:dEdx}, the shape of the \dEdx distributions in the CR and the SR agree well.

\begin{figure}[!htb]
  \centering
  \includegraphics[width=\cmsFigWidth]{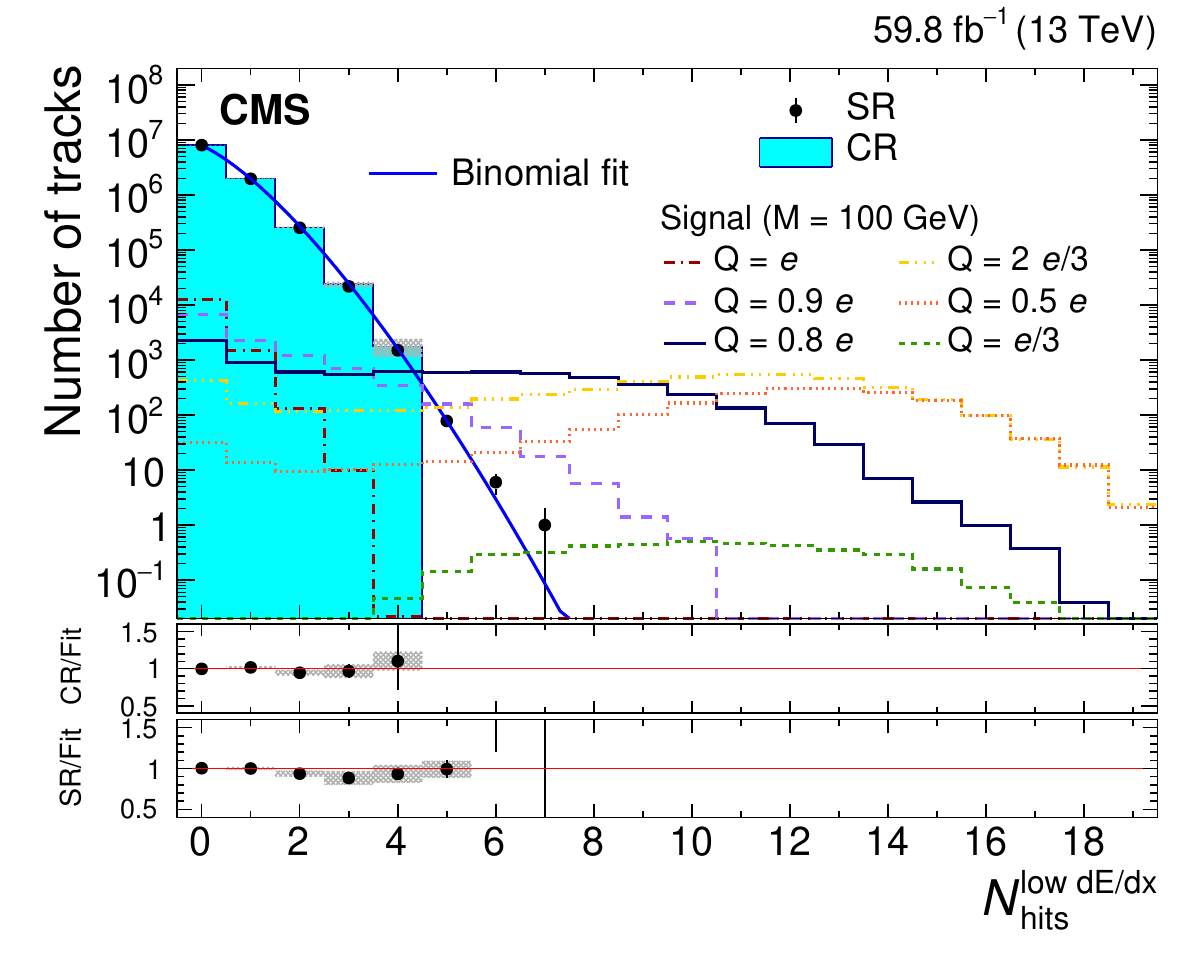}
  \caption{Distribution of \nLow in the SR and the CR for the 2018 data set, as well as for an FCP signal at a mass of 100\GeV and different charge scenarios. The vertical bars and the shaded area correspond to the statistical uncertainty in the SR and the CR, respectively. The $p$-value of the fit is 9\%. The two lower panels show the ratio of the number of tracks observed in the CR (upper) and SR (lower), and the fit function. The vertical bars correspond to the uncertainty from statistical sources, while the shaded area shows the systematic uncertainty in the binomial fit (see text).\label{fig:bgFit}}
\end{figure}

To estimate the background directly from data, we extrapolate the number of tracks in the CR from low \nLow to high \nLow values, where we expect few events.
The distribution of this variable is shown in Fig.~\ref{fig:bgFit} for 2018 data in the CR and the SR, as well as for signal events.
Corresponding figures for the other data periods can be found in \suppMaterial.
Tabulated results are provided in the HEPData record for this analysis~\cite{hepdata}.
The extrapolation is done by fitting the \nLow distribution in the CR with a binomial function.
This background evaluation method is valid for FCPs with masses larger than 50\GeV, which would give a negligible contribution to the CR.
The use of the binomial function to describe the \nLow distribution is justified by the track hits being a priori uncorrelated binary measurements by identical detector modules.
The number of binomial measurements is given by the mean number of hits per track and the probability for a measurement to have a positive outcome is given by the mean fraction of low-\dEdx hits per track.
However, instrumental effects, in particular radiation damage, affect the tracker modules in a way that the assumption that measurements come from identical detectors is not strictly met.
As a consequence, the choice of a single \dEdx threshold yields a \nLow distribution that deviates from a binomial distribution.
To restore a quasi-binomial behavior, layer-dependent upper \dEdx thresholds are chosen to enforce an equal average binomial probability for each tracker layer.
These upper thresholds typically vary around 2\MeV/\cm, below the peak of the \dEdx distribution for unit charge particles (see Fig.~\ref{fig:dEdx}).
Together with the lower threshold at 0.25\MeV/\cm, they define low-\dEdx windows in which \nLow is counted.
A $\chi^{2}$ test and the associated $p$-value~\cite{Gross:2010qma} are used to evaluate the hypothesis of the binomial distribution.
The validity of the background estimation in the SR from a fit to the CR is supported by the good agreement between the SR and the CR in the distributions of \dEdx.

In order to extract signal, we count the number of observed and expected tracks in the bins of the \nLow spectrum with the highest expected signal significance, separately for each data period.
These bins correspond to ${\nLow = 5, 6, 7, 8}$, and $\ge9$ for the early 2016 data set, and ${\nLow = 4, 5, 6}$, and $\ge7$, for the late 2016, 2017, and 2018 data sets.
In each data period, both the shape and the normalization information are used in the signal extraction.

We consider three distinct systematic uncertainties for the background prediction.
The first uncertainty arises from the fit function.
Although binomial from first principles and corrected for instrumental effects, residual correlations make the \nLow distribution slightly nonbinomial.
To account for this, we fit the distribution with an alternative function, ${N_{\mathrm{tracks}} (n) = p_{0}n^{p_{1}+p_{2}n}}$, where ${n = \nLow}$ and $p_0$, $p_1$, and $p_2$ are free parameters.
We then take the uncertainty as the relative difference between the predictions of this fit and those of the binomial function.
The second uncertainty relates to the choice of the fit range.
It is taken as the relative difference between the predictions with the binomial fit starting at ${\nLow = 0}$ and ${\nLow = 1}$, except for early 2016 data where the \nLow distribution is wider and we compare the fits starting at ${\nLow = 0}$ and ${\nLow = 2}$.
Finally, the uncertainty for the background prediction arising from the limited number of tracks in the fitted CR is negligible.

The main systematic uncertainties in the signal expectation come from the estimation of the Level-1 (L1) trigger inefficiency resulting from the large FCP mass (hence small velocity) and from the simulation of the loss of hits in the muon detector at low charge, estimated by varying dedicated muon simulation parameters.
The low FCP \dEdx causes a reconstruction inefficiency in the muon detector, for which we evaluate the uncertainty as well.
Other contributions arise from the knowledge of the integrated luminosity in the considered data sets~\cite{CMS:2021xjt,CMS-PAS-LUM-17-004,CMS-PAS-LUM-18-002} and the understanding in simulation of other event and candidate track selection requirements, in particular the track isolation and cosmic rejection.
To evaluate the uncertainty related to the correction of the underestimated fraction of low-\dEdx hits in simulation, we define \nLowQSq as \nLow with the low-\dEdx window boundaries divided by $Q^{2}$.
As $Q < e$ for FCPs, this means that we include higher \dEdx values in the window from which the hits contributing to \nLow are counted.
This reflects the fact that for a signal with low charge, the hits selected to count \nLow originate from a higher part of the \dEdx spectrum than for an equivalent unity charge particle.
The uncertainty in the signal is then obtained from the relative difference of \nLowQSq between data and DY simulation in the CR.

Table~\ref{tab:syst} shows the number of tracks expected from the background and observed for the considered data periods, for the upper inclusive \nLow bin, which has the highest sensitivity over most of the explored parameter space.
In more detailed tables that can be found in \suppMaterial, we report the systematic uncertainties for data, as well as the expected signal rate for ${Q = 2e/3}$ and $e/3$ along with the associated uncertainties.

\begin{table*}[!htb]
    \centering
    \topcaption{Number of tracks observed and expected from the background in the \nLow bin with highest signal significance for early 2016 ($\geq9$ hits), late 2016 ($\geq7$ hits), 2017 ($\geq7$ hits), and 2018 data ($\geq7$ hits). The corresponding systematic uncertainties in the background estimation are also listed.}
    \label{tab:syst}
    \begin{scotch*}{lcccc}
                                                    & early 2016  & late 2016   & 2017        & 2018      \\[\cmsTabSkip]
$N^{\text{observed}}$                               & 0           & 1           & 0           & 1         \\[\cmsTabSkip]
$N_{\text{BG}}^{\text{expected}}$ \rule{0pt}{10pt}  & 0.04        & 0.09        & 0.02        & 0.09      \\
Fit function                                        & $\pm 290$\% & $\pm 160$\% & $\pm 230$\% & $\pm 19$\%\\
Fit range                                           & $\pm 110$\% & $\pm 0.3$\% & $\pm 14$\%& $\pm 28$\%\\
    \end{scotch*}
\end{table*}

\begin{figure}[!htb]
  \centering
  \includegraphics[width=\cmsFigWidth]{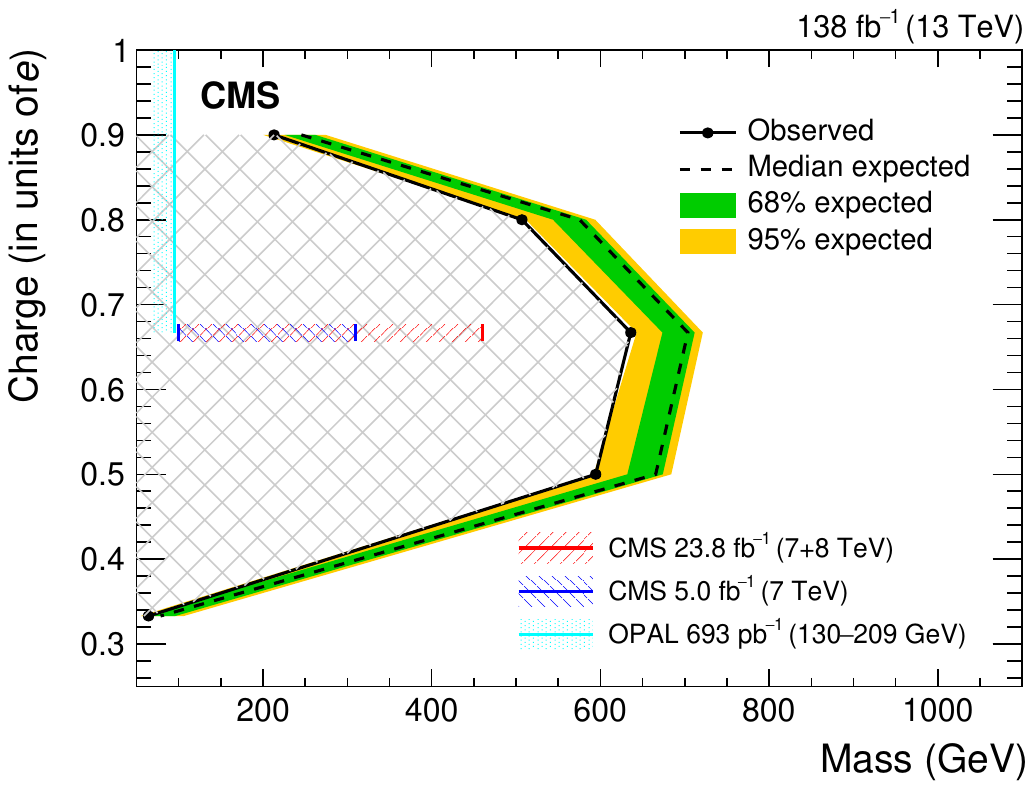}
  \caption{Exclusion region (hatched) at 95\% \CL in the FCP charge-mass plane for the considered signal. The expected exclusion is shown with the associated 1 (green) and 2 (orange) standard deviations bands. Signal points at charges 0.9, 0.8, 2/3, 0.5, and $1/3e$ are connected by straight lines to guide the eye. This is a conservative interpolation. Previous exclusions from CMS~\cite{CMS:2012xi,CMS:2013czn} as well as OPAL~\cite{OPAL:2003zpa} are given for comparison.\label{fig:limits}}
\end{figure}

No significant deviation is observed when comparing data with the background prediction in all search region bins across all considered data sets.
Using the \CLs method~\cite{Junk:1999kv,Read:2002hq}, 95\% CL upper limits are computed for the FCP production cross section for a given charge and mass scenario.
The excluded parameter space for the considered signal is shown in Fig.~\ref{fig:limits}, using the full 2016--2018 data set.
The exclusion shown is valid for masses above 50\GeV.
The limit for a charge of $2e/3$ is improved with respect to prior analyses, and we extend the exclusion to other charge points, from $0.9e$ down to $e/3$.
For an FCP of charge $2e/3$ ($e/3$), we exclude cross sections above 0.28\unit{fb} (0.39\unit{pb}) at the maximal excluded mass of 640 (60)\GeV.

In summary, we presented a search for fractionally charged particles using proton-proton collisions corresponding to an integrated luminosity of 138\fbinv, collected at ${\sqrt{s} = 13\TeV}$ with the CMS detector.
The key feature of low ionization energy in the CMS tracker detector is used to discriminate a potential signal from the background.
No significant deviation is observed in data with respect to the expected background.
The search presents for the first time a scan of the new particle charge between $e/3$ and $0.9e$ at the LHC.
The existence of fractionally charged particles arising from the considered DY-like pair production is excluded in the mass range between 50 and 640 (60)\GeV for a signal of charge ${Q = 2e/3}$ ($e/3$).
These are the most stringent limits to date for this production mode.

\begin{acknowledgments}
We congratulate our colleagues in the CERN accelerator departments for the excellent performance of the LHC and thank the technical and administrative staffs at CERN and at other CMS institutes for their contributions to the success of the CMS effort. In addition, we gratefully acknowledge the computing centres and personnel of the Worldwide LHC Computing Grid and other centres for delivering so effectively the computing infrastructure essential to our analyses. Finally, we acknowledge the enduring support for the construction and operation of the LHC, the CMS detector, and the supporting computing infrastructure provided by the following funding agencies: SC (Armenia), BMBWF and FWF (Austria); FNRS and FWO (Belgium); CNPq, CAPES, FAPERJ, FAPERGS, and FAPESP (Brazil); MES and BNSF (Bulgaria); CERN; CAS, MoST, and NSFC (China); MINCIENCIAS (Colombia); MSES and CSF (Croatia); RIF (Cyprus); SENESCYT (Ecuador); ERC PRG, RVTT3 and TK202 (Estonia); Academy of Finland, MEC, and HIP (Finland); CEA and CNRS/IN2P3 (France); SRNSF (Georgia); BMBF, DFG, and HGF (Germany); GSRI (Greece); NKFIH (Hungary); DAE and DST (India); IPM (Iran); SFI (Ireland); INFN (Italy); MSIP and NRF (Republic of Korea); MES (Latvia); LAS (Lithuania); MOE and UM (Malaysia); BUAP, CINVESTAV, CONACYT, LNS, SEP, and UASLP-FAI (Mexico); MOS (Montenegro); MBIE (New Zealand); PAEC (Pakistan); MES and NSC (Poland); FCT (Portugal); MESTD (Serbia); MCIN/AEI and PCTI (Spain); MOSTR (Sri Lanka); Swiss Funding Agencies (Switzerland); MST (Taipei); MHESI and NSTDA (Thailand); TUBITAK and TENMAK (Turkey); NASU (Ukraine); STFC (United Kingdom); DOE and NSF (USA).  
\end{acknowledgments}

\bibliography{auto_generated}

\ifthenelse{\boolean{cms@external}}{}{
\clearpage
\numberwithin{table}{section}
\numberwithin{figure}{section}
\appendix
\section{Additional figures and tables\label{app:suppMat}}
\newlength\cmsFigWidthii\setlength\cmsFigWidthii{0.5\textwidth}
\begin{figure}[htpb!]
  \centering
  \includegraphics[width=\cmsFigWidthii]{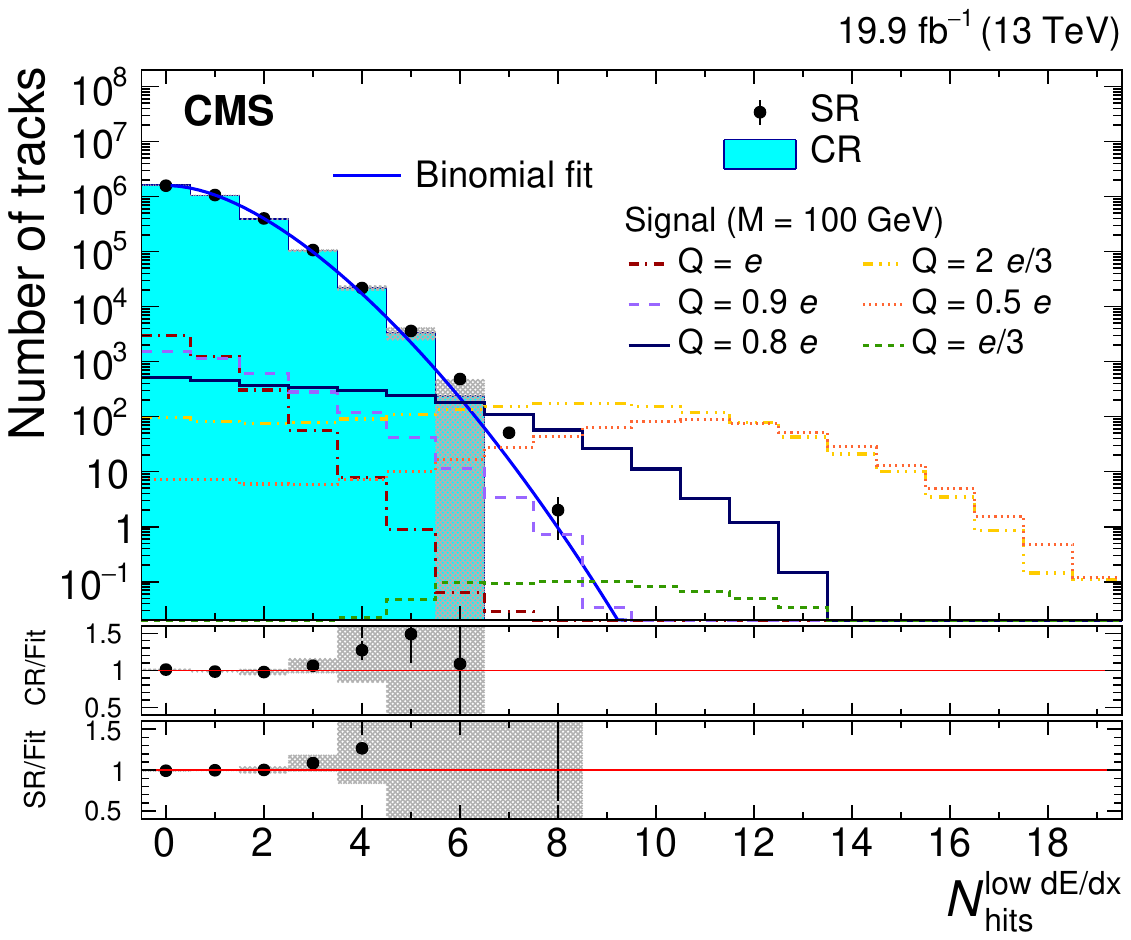}
  \caption{Distribution of \nLow in the SR and the CR for the early 2016 data set, as well as for an FCP signal at a mass of 100\GeV and different charge scenarios. The vertical bars and the shaded area correspond to the statistical uncertainty in the SR and the CR, respectively. The $p$-value of the fit is 6\%. The two lower panels show the ratio of the number of tracks observed in the CR (upper) and SR (lower), and the fit function. The vertical bars correspond to the uncertainty from statistical sources, while the shaded area shows the systematic uncertainty in the fit due to the choice of the fitting function and the binomial fit range as explained in the main text. Comparing with respect to the binomial fit starting at $\nLow = 2$, and not $\nLow = 1$, is needed to account for the fact that early 2016 data is more strongly affected by instrumental effects that widen the \nLow distribution.}
\end{figure}

\begin{figure}[htpb!]
  \centering
  \includegraphics[width=\cmsFigWidthii]{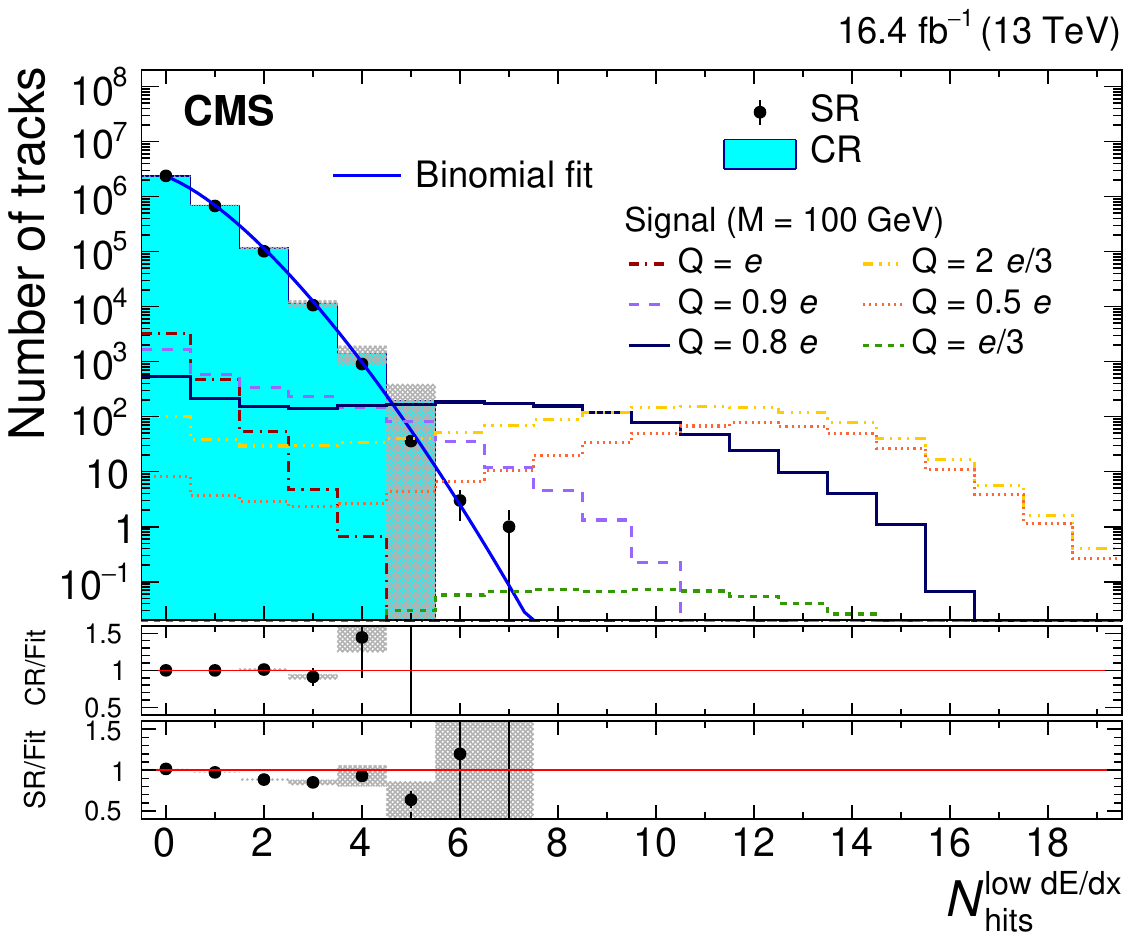}
  \caption{Distribution of \nLow in the SR and the CR for the late 2016 data set, as well as for an FCP signal at a mass of 100\GeV and different charge scenarios. The vertical bars and the shaded area correspond to the statistical uncertainty in the SR and the CR, respectively. The $p$-value of the fit is 78\%. The two lower panels show the ratio of the number of tracks observed in the CR (upper) and SR (lower), and the fit function. The vertical bars correspond to the uncertainty from statistical sources, while the shaded area shows the systematic uncertainty in the fit due to the choice of the fitting function and the binomial fit range as explained in the main text.}
\end{figure}

\clearpage
\begin{figure}[htpb!]
  \centering
  \includegraphics[width=\cmsFigWidthii]{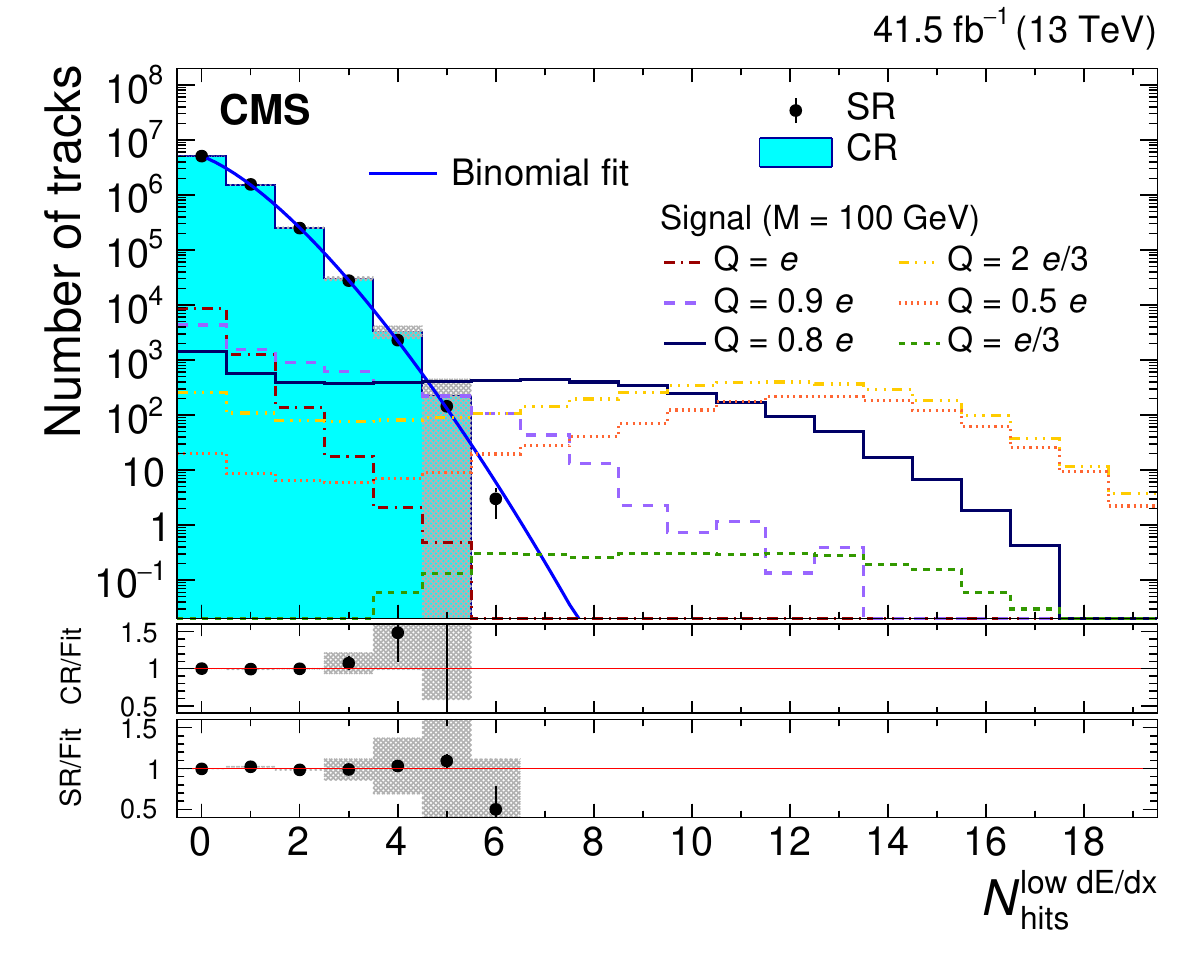}
  \caption{Distribution of \nLow in the SR and the CR for the 2017 data set, as well as for an FCP signal at a mass of 100\GeV and different charge scenarios. The vertical bars and the shaded area correspond to the statistical uncertainty on the SR and the CR, respectively. The $p$-value of the fit is 65\%. The two lower panels show the ratio of the number of tracks observed in the CR (upper) and SR (lower), and the fit function. The vertical bars correspond to the uncertainty from statistical sources, while the shaded area shows the systematic uncertainty in the fit due to the choice of the fitting function and the binomial fit range as explained in the main text.}
\end{figure}

\begin{table*}[htpb!]
  \centering
    \topcaption{Number of tracks observed and expected for background and signal events, in the \nLow bin with highest signal significance ($\geq$9), for early 2016 data. The corresponding systematic uncertainties are also listed. They are considered to follow a log-normal distribution. The results are shown for FCP scenarios at a mass of 100\GeV with a charge of 2$e$/3 (the most sensitive scenario) and $e$/3 (the lowest considered charge).}
    \begin{scotch*}{lcc}
$N^{\text{observed}}$ \rule{0pt}{10pt}              & \multicolumn{2}{c}{0}\\[\cmsTabSkip]
$N_{\text{BG}}^{\text{expected}}$ \rule{0pt}{10pt}  & \multicolumn{2}{c}{0.04}\\
Fit function                                        & \multicolumn{2}{c}{$\pm 287$\%}\\
Fit range                                           & \multicolumn{2}{c}{$\pm 109$\%}\\[\cmsTabSkip]
\rule{0pt}{10pt}                                               & $Q = 2e/3$ & $Q = e/3$\\
                                                               \cline{2-3}
$N_{\text{signal}}^{\text{expected}}$ \rule{0pt}{10pt}  & 602   & 0.328\\
L1 trigger inefficiency                                        & 5\%   & 5\%\\
Muon system energy loss simul.                                 & $+$4\%/$-$3\%    & 75\%\\
Muon detector inefficiency                                     & 3\%   & 15\%\\
Integrated luminosity                                          & 1.2\% & 1.2\%\\
Selection                                                      & 2.0\% & 2.0\%\\
\dEdx corrections                                              & 3.8\% & 12.2\%\\
Limited signal sample                                          & $<$1\% & 2.4\%\\
    \end{scotch*}
\end{table*}

\begin{table}[htpb!]
  \centering
    \topcaption{Number of tracks observed and expected for background and signal events, in the \nLow bin with highest signal significance ($\geq$7), for late 2016 data. The corresponding systematic uncertainties are also listed. They are considered to follow a log-normal distribution. The results are shown for FCP scenarios at a mass of 100\GeV with a charge of $2e/3$ (the most sensitive scenario) and $e/3$ (the lowest considered charge).}
    \begin{scotch*}{lcc}
$N^{\text{observed}}$ \rule{0pt}{10pt}              & \multicolumn{2}{c}{1}\\[\cmsTabSkip]
$N_{\text{BG}}^{\text{expected}}$ \rule{0pt}{10pt}  & \multicolumn{2}{c}{0.09}\\
Fit function                                        & \multicolumn{2}{c}{$\pm 159$\%}\\
Fit range                                           & \multicolumn{2}{c}{$\pm 0.3$\%}\\[\cmsTabSkip]
\rule{0pt}{10pt}                                               & $Q = 2e/3$ & $Q = e/3$\\
                                                               \cline{2-3}
$N_{\text{signal}}^{\text{expected}}$ \rule{0pt}{10pt}  & 988 & 0.454 \\
L1 trigger inefficiency                                        & 5\%   & 5\%\\
Muon system energy loss simul.                                 & $+$4\%/$-$3\%    & 75\%\\
Muon detector inefficiency                                     & 3\%   & 15\%\\
Integrated luminosity                                          & 1.2\% & 1.2\%\\
Selection                                                      & 2.0\% & 2.0\%\\
\dEdx corrections                                              & 1.1\% & 1.4\%\\
Limited signal sample                                          & $<$1\% & 2.6\%\\
    \end{scotch*}
\end{table}

\begin{table}[htpb!]
  \centering
    \topcaption{Number of tracks observed and expected for background and signal events, in the \nLow bin with highest signal significance ($\geq$7), for 2017 data. The corresponding systematic uncertainties are also listed. They are considered to follow a log-normal distribution. The results are shown for FCP scenarios at a mass of 100\GeV with a charge of $2e/3$ (the most sensitive scenario) and $e/3$ (the lowest considered charge).}
    \begin{scotch*}{lcc}
$N^{\text{observed}}$ \rule{0pt}{10pt}              & \multicolumn{2}{c}{0}\\[\cmsTabSkip]
$N_{\text{BG}}^{\text{expected}}$ \rule{0pt}{10pt}  & \multicolumn{2}{c}{0.2}\\
Fit function                                        & \multicolumn{2}{c}{$\pm 231$\%}\\
Fit range                                           & \multicolumn{2}{c}{$\pm 13.7$\%}\\[\cmsTabSkip]
\rule{0pt}{10pt}                                               & $Q = 2e/3$ & $Q = e/3$\\
                                                               \cline{2-3}
$N_{\text{signal}}^{\text{expected}}$ \rule{0pt}{10pt}  & 2720 & 2.25 \\
L1 trigger inefficiency                                        & 5\%   & 5\%\\
Muon system energy loss simul.                                 & $+$4\%/$-$3\%    & 75\%\\
Muon detector inefficiency                                     & 3\%   & 15\%\\
Integrated luminosity                                          & 2.3\% & 2.3\%\\
Selection                                                      & 2.0\% & 2.0\%\\
\dEdx corrections                                              & 2.3\% & 3.3\%\\
Limited signal sample                                          & $<$1\% & 1.5\%\\
    \end{scotch*}
\end{table}

\begin{table*}[!htb]
    \centering
    \topcaption{Number of tracks observed and expected for background and signal events, in the \nLow bin with highest signal significance ($\geq$7), for 2018 data. The corresponding systematic uncertainties are also listed. They are considered to follow a log-normal distribution. The results are shown for FCP scenarios at a mass of 100\GeV with a charge of $2e/3$ (the most sensitive scenario) and $e/3$ (the lowest considered charge).}
    \begin{scotch*}{lcc}
$N^{\text{observed}}$ \rule{0pt}{10pt}              & \multicolumn{2}{c}{1}\\[\cmsTabSkip]
$N_{\text{BG}}^{\text{expected}}$ \rule{0pt}{10pt}  & \multicolumn{2}{c}{0.09}\\
Fit function                                        & \multicolumn{2}{c}{$\pm 19$\%}\\
Fit range                                           & \multicolumn{2}{c}{$\pm 28$\%}\\[\cmsTabSkip]
\rule{0pt}{10pt}                                               & $Q = 2 e/3$ & $Q = e/3$\\
                                                               \cline{2-3}
$N_{\text{signal}}^{\text{expected}}$ \rule{0pt}{10pt}  & 3622  & 3.51\\
L1 trigger inefficiency                                        & 5\%    & 5\%\\
Muon system energy loss simul.                                 & $+$4\%/$-$3\%    & 75\%\\
Muon detector inefficiency                                     & 3\%    & 15\%\\
Integrated luminosity                                          & 2.5\%  & 2.5\%\\
Selection                                                      & 2.0\%  & 2.0\%\\
\dEdx corrections                                              & 1.7\%  & 2.3\%\\
Limited signal sample                                          & $<$1\% & 1.8\%\\
    \end{scotch*}
\end{table*}

}
\cleardoublepage \section{The CMS Collaboration \label{app:collab}}\begin{sloppypar}\hyphenpenalty=5000\widowpenalty=500\clubpenalty=5000
\cmsinstitute{Yerevan Physics Institute, Yerevan, Armenia}
{\tolerance=6000
A.~Hayrapetyan, A.~Tumasyan\cmsAuthorMark{1}\cmsorcid{0009-0000-0684-6742}
\par}
\cmsinstitute{Institut f\"{u}r Hochenergiephysik, Vienna, Austria}
{\tolerance=6000
W.~Adam\cmsorcid{0000-0001-9099-4341}, J.W.~Andrejkovic, T.~Bergauer\cmsorcid{0000-0002-5786-0293}, S.~Chatterjee\cmsorcid{0000-0003-2660-0349}, K.~Damanakis\cmsorcid{0000-0001-5389-2872}, M.~Dragicevic\cmsorcid{0000-0003-1967-6783}, A.~Escalante~Del~Valle\cmsorcid{0000-0002-9702-6359}, P.S.~Hussain\cmsorcid{0000-0002-4825-5278}, M.~Jeitler\cmsAuthorMark{2}\cmsorcid{0000-0002-5141-9560}, N.~Krammer\cmsorcid{0000-0002-0548-0985}, L.~Lechner\cmsorcid{0000-0002-3065-1141}, D.~Liko\cmsorcid{0000-0002-3380-473X}, I.~Mikulec\cmsorcid{0000-0003-0385-2746}, J.~Schieck\cmsAuthorMark{2}\cmsorcid{0000-0002-1058-8093}, R.~Sch\"{o}fbeck\cmsorcid{0000-0002-2332-8784}, D.~Schwarz\cmsorcid{0000-0002-3821-7331}, M.~Sonawane\cmsorcid{0000-0003-0510-7010}, S.~Templ\cmsorcid{0000-0003-3137-5692}, W.~Waltenberger\cmsorcid{0000-0002-6215-7228}, C.-E.~Wulz\cmsAuthorMark{2}\cmsorcid{0000-0001-9226-5812}
% ...
\par}
\vskip\cmsinstskip
\dag:~Deceased\\
$^{1}$Also at Yerevan State University, Yerevan, Armenia\\
$^{2}$Also at TU Wien, Vienna, Austria\\
% ...
\end{sloppypar}
\end{document}